\documentclass[twocolumn,showpacs,preprintnumbers,amsmath,amssymb]{revtex4}
\usepackage{amssymb}
\usepackage[dvips]{graphicx}
\begin{document}

\title{
The "normal" state of superconducting cuprates might really be normal after all
}
\author{V. N. Zavaritsky$^{1,2}$ and A. S. Alexandrov$^{1}$}

\address
{$^{1}$Loughborough University, Loughborough, United Kingdom\\
$^{2}$Kapitza Institute for Physical Problems \& General Physics Institute,  Moscow, Russia\\
}

\begin{abstract}
High magnetic field studies of  cuprate superconductors revealed    a non-BCS  temperature
 dependence of the upper critical field $H_{c2}(T)$ determined resistively by several groups.
 These determinations caused some doubts on the grounds of both the
 contrasting effect of the magnetic field on the in-plane and out-of-plane
 resistances reported for large Bi2212 sample and the large Nernst signal \emph{well above}  $T_{c}$.
 Here we present both $\rho_{ab}(B)$ and $\rho_{c}(B)$ of tiny Bi2212 crystals in  magnetic fields up to 50 Tesla.
 None of our measurements revealed a situation when on the field increase $\rho_c$ reaches
 its maximum while $\rho_{ab}$ remains very small if not zero.
 The resistive %upper critical fields estimated from the in-plane and out-of-plane
$H_{c2}(T)$ estimated from $\rho_{ab}(B)$ and $\rho_{c}(B)$ are
approximately the same.  Our results  support any theory of
cuprates that describes the state above the resistive phase
transition as perfectly normal with a zero off-diagonal order
parameter. In particular, the anomalous Nernst effect above the resistive phase transition in
high-$T_{c}$ cuprates can be  described quantitatively as a normal state phenomenon in a  model with
itinerant and localised fermions and/or charged bosons.
\end{abstract}

\pacs{74.40.+k, 72.15.Jf, 74.72.-h, 74.25.Fy}

\maketitle

A  pseudogap is believed to be responsible for the non Fermi-liquid normal state of cuprate superconductors. Various
microscopic models of the pseudogap proposed are mostly based on strong electron correlations \cite{tim}, and/or  on  strong electron-phonon interaction\cite{alebook}. There is also a phenomenological scenario \cite{Kiv}, where the superconducting order parameter (the Bogoliubov-Gor'kov anomalous average $F(\mathbf{r,r^{\prime}})=\langle \psi _{\downarrow}(\mathbf{{r})\psi _{\uparrow}({r^{\prime }}\rangle }$) does not disappear at %the resistive 
$T_{c}$ but at much higher (pseudogap) temperature. While the scenario \cite{Kiv} was found to be inconsistent with the  `intrinsic tunnelling' I-V characteristics, the discovery of the Joule heating origin  of the gap-like I-V nonlinearities made that objection irrelevant \cite{ije-phys}. Some other measurements \cite{mein-cor} also provide evidence in support of \cite{Kiv}.

In line with the scenario, several authors \cite{mor,xu} suggested a radical revision of the magnetic phase diagram of the cuprates with an upper critical field much higher than the resistive $H_{c2}(T)$. In particular,  Ref.\cite{mor} questioned the resistive determination of $H_{c2}(T)$ \cite{alezav,zav}, claiming that, while $\rho_c$ measures the inter-plane tunnelling, only the in-plane data represent a true normal state. The main argument in favour of this claim came from the radically different field dependencies of $\rho_c$ and $\rho_{ab}$ in Ref.\cite{mor} (shown below in  our Fig.2B). According to this finding, a magnetic field sufficient to recover the normal state $\rho_c$, leaves in-plane superconductivity virtually unaffected. This discrepancy suggests that Bi2212 crystals do not lose their  off-diagonal order in Cu0$_2$ planes even well above $H_{c2}(T)$ determined from $\rho_c(B,T)$. However, this conclusion is based on one measurement and so certainly deserves experimental verification, which was not possible until recently because of the lack of reliable $\rho_{ab}(B,T)$ for Bi2212. 

Quite similar conclusions followed from thermomagnetic studies of superconducting cuprates. Here a large Nernst signal  \emph{well above}  $T_{c}$  has been attributed to a \emph{vortex} motion. As a result, the magnetic phase diagram of the cuprates has been revised radically. Most surprisingly, Ref.\cite{xu} estimated $H_{c2}$ \emph {at the zero-field transition temperature}, $T_{c0}$, of Bi2212 as high as 50-150\,Tesla.

On the other hand, any  scenario  with  $F(\mathbf{r,r^{\prime}})\neq0$  in the "normal" state  is difficult to reconcile with the extremely sharp resistive and magnetic transitions at $T_{c}$ in single crystals of cuprates. Above $T_c$, the uniform magnetic susceptibility is paramagnetic and the resistivity is perfectly 'normal',  showing only a few percent positive or negative magnetoresistance (MR). Both in-plane \cite{mac,boz,fra}  and out-of-plane \cite{alezav} resistive transitions remain sharp in the magnetic field in  high-quality samples, providing a reliable determination of a genuine $H_{c2}(T)$.   These and some other observations \cite{lor}  do not support any  \emph{stationary} superconducting order parameter above $T_{c}$.

Resolution of these issues, which affect fundamental conclusions about the nature of superconductivity in highly anisotropic layered cuprates, requires further careful experiments and transparent interpretations. Here we present systematic measurements of both in-plane and out-of-plane MRs of small  Bi2212 single crystals subjected to magnetic fields, $B\leq50$ Tesla, $B\perp(ab)$. Our measurements reproduced neither the unusual field dependence of $\rho_{ab}$  nor the contrasting effect of the field as in  \cite{mor}, which are most probably an experimental artefact. On the contrary, they show that $H_{c2}(T)$ estimated from $\rho_{ab}$ and $\rho_c$ are nearly identical. These results, along with a simple explanation of the unusual Nernst signal in cuprates as a normal state phenomenon \cite{NERNST}, strongly support any microscopic theory of cuprates with a zero off-diagonal order parameter above resistive $T_c(B)$.

\begin{figure}
\begin{center}
\includegraphics[angle=-0,width=0.47\textwidth]{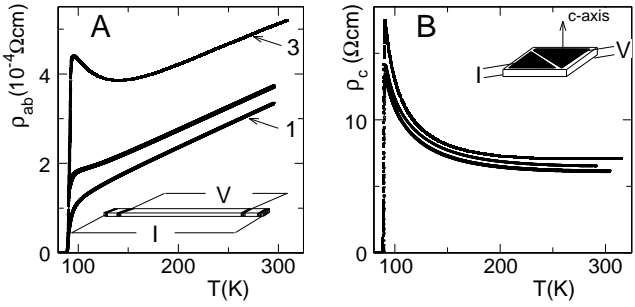}
\vskip -0.5mm \caption{Contact layout and examples of $\rho_c(T)$ and $\rho_{ab}(T)$ measured on small samples cleaved from the same Bi2212 crystal. $\rho_c$ contamination of $\rho_{ab}$ is $\sim10^{-5}$ and indistinguishable   for the curves labelled as 3 and 1 respectively.}
\end{center}
\end{figure}

Reliable measurements of the resistivity tensor require defect-free samples.  This is of prime importance for in-plane MR because, owing to the extreme anisotropy of Bi2212 \cite{mos}, even unit-cell scale  defects will result in a significant out-of-plane contribution. Not only are such minor defects impossible to detect by conventional techniques, but  $\rho_{ab}$ contamination with $\rho_c$ might occur even in a perfect crystal with nonuniform current distribution. For these reasons, we paid special attention to  sample preparation and selection \cite{mos}. Since the extremely high and temperature dependent electric anisotropy of Bi2212 prevents reliable measurement of both the in-plane and out-of-plane resistances on the same sample, we measured $\rho_{c}$ and $\rho_{ab}$ on different pieces of the same high-quality, optimally and slightly underdoped Bi2212 parent crystals with T$_{c0}\approx$87-92$K$. As the specific demands of pulsed field experiments make it essential to use tiny specimens, we measured $\rho_{c}$ on samples with in-plane dimensions from $\simeq 30 \times 30 \mu m^2$ to $\simeq 80 \times 80 \mu m^2$, while $\rho_{ab}$ was studied on  longer crystals, from $\simeq300\times11\mu m^2$ to $\simeq780\times22\mu m^2$. The samples for this study were selected on the basis of comparative analysis of transport measurements of 7-12 pairs of such samples, cleaved from  different places of the same parent crystal (typically of $1-3\mu m$ thickness). To achieve a uniform {\it in-plane} current distribution, the current contacts were made by immersion of the crystals' ends into diluted conductive composite; $\rho_c$ was measured with the contacts deposited on both ab-faces, see Fig.1. The uncertainty of the samples' dimensions is most probable cause of the mismatch of $\rho_c$ in different pieces, Fig.1B. Unlike $\rho_c(T)$ curves, $\rho_{ab}(T)$ of different pieces often reveal qualitatively different behaviour, illustrated in Fig.1A. While the majority of the `$\rho_{ab}$-samples' had the metallic type of zero-field $\rho_{ab}(T)$ represented by the curve 1, others demonstrated the sample-dependent $\rho_{ab}(T)$ upturn, which we attribute to $\rho_c$ contamination. Only the samples with the lowest $\rho_{ab}(T)$ were selected for this study. The metallic type of zero-field $\rho_{ab}(T)$ and the {\it sign} of its normal state MR \cite{mos} indicate a vanishing $\rho_{c}$-contribution. The absence of hysteresis in the $\rho(B)$ data obtained on the rising and falling sides of the pulse and the consistency of $\rho(B)$ taken at the same temperature in pulses of different $B_{max}$ exclude any measurable heating effects. The Ohmic response is confirmed by the consistency of $\rho(B)$ measured at identical conditions with different currents, $10$-$1000A/cm^2$ for $\rho_{ab}$ and $0.1$-$20A/cm^2$ for $\rho_c$.

\begin{figure}
\begin{center}
\includegraphics[angle=-0,width=0.47\textwidth]{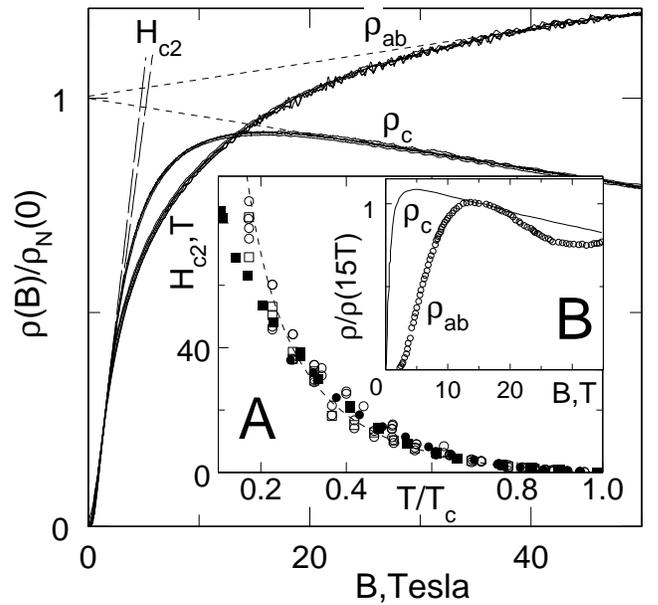}
\vskip -0.5mm \caption{ $\rho_c(B)$ and $\rho_{ab}(B)$ of Bi2212 at $\sim$68K, normalised by corresponding $\rho_N(0,T)$ obtained with the linear extrapolation from the normal state region (short dashes). The linear fits, shown by long dashed lines, refer to the flux-flow region. Inset A:   $H_{c2}$ estimated  from $\rho_{ab}(B)$ and $\rho_{c}(B)$  is shown by the open and  solid symbols respectively together with the fit, $H_{c2}(T) \sim (t^{-1}-t^{1/2})^{3/2}$, with $t=T/T_{c}$ \cite{ale} (broken line). Inset B shows $\rho_c$ and $\rho_{ab}$ from  Ref.\cite{mor}.
}
\end{center}
\end{figure}
Fig.2 shows the typical $\rho_c(B)$ and $\rho_{ab}(B)$  taken below $T_{c0}$ of a Bi2212 single crystal. The low-field portions of the curves correspond to the resistance driven by vortex dynamics. Here, a non-linear $\rho(B)$ dependence is followed by a regime in which linear dependence fits the experimental observations rather well, Fig.2.  It is natural to attribute the high field portions of the curves in Fig.2 (assumed to be above H$_{c2}$) to a normal state \cite{zav}. Here, the c-axis high-field MR appears to be negative and quasi-linear in B in a wide temperature range both above and below $T_{c0}$. Contrary to  $\rho_{c}(B)$, the normal state in-plane MR is {\it positive} (see \cite{mos} and references therein for   an explanation). The resistive upper critical field, H$_{c2}$(T), is estimated from $\rho_c(B)$ and $\rho_{ab}(B)$ either as the intersection of two linear approximations in Fig.2, or from the  flux-flow resistance as $H_{c2}=\rho_N(0,T)(\partial \rho_{FF}/\partial B)^{-1}$; both estimates are found to be almost identical. This procedure allows us to separate contributions \mbox{originating} from the normal and superconducting states and, in particular, to avoid ambiguity resulting from fluctuations in the crossover region. The downward deviations from the linear field dependence at fields around $H_{c2}$ in Fig.2 are most likely caused by the conventional (3D-XY \cite{tony}) critical behaviour rather than the stationary off-diagonal order parameter in the "normal" phase \cite{tau}. The reasonable concordance of $H_{c2}(T)$ estimates from $\rho_{c}(B)$ and $\rho_{ab}(B)$ (Fig.2A) favours our association of the resistive $H_{c2}$ with the upper critical field, especially given the apparently different mechanisms responsible for $\rho_{ab}$ and $\rho_{c}$ \cite{mos}.

Our conclusion is based on the results obtained during several hundred measurements performed on three pairs of crystals. None of those revealed a situation in which on field increase $\rho_c$ reaches its
maximum while $\rho_{ab}$ remains very small if not zero as in \cite{mor} (see Fig.2B). Since the authors of Ref.\cite{mor}  measured $'\rho_{ab} (B)'$ by means of contacts situated on the same face of the crystal while the current was injected into the opposite face, their curve could {\it not} represent the true $\rho_{ab}$. We cannot exclude the possibility that this observation might be caused by current redistribution in the medium with field and temperature dependent anisotropy. This opinion is supported by the independent study of current redistribution  in homogeneous  Bi2212, \cite{bush}. However, the threefold $\rho_c$ enhancement warrants inhomogeneity of the huge crystal in \cite{mor} so that the results of \cite{bush} may not be directly applicable to this case.  Neither the current redistribution nor imperfections of the crystal were accounted for in \cite{mor}.

\begin{figure}
\begin{center}
\includegraphics[angle=-0,width=0.47\textwidth]{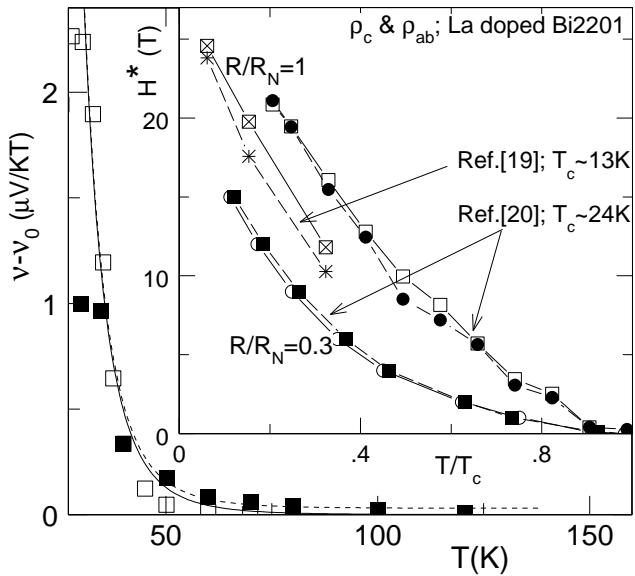}
\vskip -0.5mm \caption{$\square$ and $\blacksquare$ represent $\nu=e_y/B$ measured on Bi2201 \cite{wang-prb} and YBCO \cite{annalen} respectively; the solid and broken lines show fit with $a\propto T^{-6}-\nu_0$ for $\nu_0$=0 and  -0.03$\mu V/KT$ respectively. Inset: $H_{c2}(T)$  obtained from the independent resistive studies of Bi2201 \cite{ando,zhang}; the broken and solid lines correspond to the data taken from $\rho_c$ and $\rho_{ab}$ respectively. }
\end{center}
\end{figure}

Our conclusions are supported by independent studies of a single-layer cuprate Bi(La)2201 with similar anisotropy.  If we apply the routine procedure for the $H_{c2}(T)$ evaluation\cite{alezav}, very similar values of $H_{c2}(T)$ are obtained from $\rho_{ab}$ $and$ $\rho_c$  measured on {\it the same} crystals \cite{ando} and films \cite{zhang} (see broken and solid lines in the inset to Fig.3). The functional similarity of $H_{c2}(T)$ dependences estimated for the same conditions from resistivities of physically different origin is evident from  Fig.2A and Fig.3(inset). Remarkably, these $H_{c2}(T)$ are compatible with the Bose-Einstein condensation field of preformed charged bosons \cite{ale} (Fig.2A), and also with some other models \cite{abr}. The described experiments  were  performed in optimally doped or only slightly underdoped samples. It would be desirable to extend these studies to more underdoped samples, where the conditions for bosonic superconductivity \cite{alebook} are definitely satisfied.

Finally, we briefly address the origin of the Nernst effect in {\it superconducting} cuprates, which is found to be enormous well above $T_c$, in drastic contrast with conventional superconductors. While a significant fraction of research in the field of high-temperature superconductivity \cite{phase} describes the unusual Nernst signal as a signature of a nonzero superconducting order parameter above (resistive) $T_c$, a key to resolution of this dichotomy lies most likely in a qualitatively different normal state of cuprates as compared with conventional superconductors. While the latter are reasonably good metals, cuprates are known to be non-stoichiometric compounds. Moreover, undoped cuprates are  insulators and their (super)conductivity appears as a result of doping, which inevitably introduces additional disorder. For these reasons, the conventional theory of heavily doped semiconductors and disordered metals might provide an adequate description of the normal state kinetic properties of cuprates (see \cite{NERNST} for more details). Carriers in doped semiconductors occupy states localised by disorder and itinerant Bloch-like states. Both types of carriers contribute to transport properties if the chemical potential $\mu$ (or the Fermi level) is close to the energy at which the lowest itinerant state appears (i.e. the mobility edge). When the chemical potential is near the mobility edge, and  the effective mass approximation is applied, there is no Nernst signal from itinerant carriers alone because of the so-called Sondheimer cancellation \cite{sond}. However, when localised carriers contribute to the longitudinal transport, a finite positive Nernst signal $ e_y\equiv -E_y/\triangledown_x T$ appears as \cite{NERNST}
\begin{equation}
\frac{e_y}{\rho}={\frac{k_{B}}{{e}}}r\theta \sigma _{l},
\end{equation}
where $\rho=1/[(2s+1)\sigma_{xx}]$ is the resistivity, $s$ is the carrier spin, $r$ is nearly constant ($r$$\approx$14.3 for fermions $s$=1/2, $r$$\approx$2.4 for bosons $s$=0), and $\Theta$ is the Hall angle. Here, $\sigma_{xx}$ is the conductivity of itinerant carriers, and $\sigma_l$ is the conductivity of localised carriers, which obeys Mott's law, $\sigma_l= \sigma_0 \exp \left[-(T_0/T)^{1/3}\right]$. In two dimensions  $T_0 \approx 8\alpha^2/(k_BN_l)$, where $N_l$ is the DOS at the Fermi level\cite{mot,shk,tok}.

In a sufficiently strong magnetic field, the radius of the 'impurity' wave function $\alpha^{-1}$ is about the magnetic length, $\alpha$$ \approx$$(eB)^{1/2}$.  Then the Nernst signal is given by
\begin{equation}
\frac{e_y}{B\rho}= a(T) \exp\left[-b(B/T)^{1/3}\right],
\end{equation}
where $a(T) \propto T^{-6}$  \cite{NERNST} and $b=2[e/(k_BN_l)]^{1/3}$ is a constant determined by the density of impurities. As follows from Eq.(2), $a(T)$ is mostly responsible for the temperature dependence of the Nernst voltage above $T_c(B)$. Interestingly, the temperature dependences of the experimental Nernst voltage taken at fixed field, $e_y/B$, in LSCO, YBCO, Bi2212, and Bi(La)2201 single crystals agree reasonably with this theoretical result, as is illustrated in Fig.3 for YBCO-0.45 and Bi(La)2201-0.4. Although this qualitative agreement favours our model,  convincing verification of the theory  would be provided by analysis  of the complementary $e_y(B,T)$ {\it and} $\rho(B,T)$ 2D-arrays of experimental data. This decisive comparison was recently performed. Remarkably, we found that the single-parameter relation, Eq.(2), {\it quantitatively} describes  both the {\it field and temperature} dependencies of $e_y/(B\rho)$ measured experimentally above the resistive critical temperature $T_c(B)$ (see Ref.\cite{NERNST} for more details). Thus we conclude that the simple model with itinerant and localised fermions and/or charged bosons is compatible with most significant thermomagnetic and kinetic measurements in superconducting cuprates. 

To conclude, we have shown that   reliable experimental data do not require radical revision of the magnetic phase diagram of cuprates \cite{zavkabale}. In particular, the reasonable concordance of resistive upper critical fields estimated from $\rho_{ab}(B)$ and $\rho_c(B)$ favours our assignment of resistive $H_{c2}$ to the genuine upper critical field, especially given the apparently different mechanisms responsible for the in-plane and out-of-plane resistivity in the normal state of Bi2212 and Bi(La)2201, as evidenced by the huge and temperature dependent anisotropy, $\rho_{c}/\rho_{ab}\geq10^4-10^5$. Our experimental  $\rho_{ab}(T,B)$ and $\rho_c(T,B)$ in the same Bi2212 crystals and the model of the Nernst signal support virtually any microscopic theory that describes the state above the resistive and magnetic phase transition in superconducting cuprates as perfectly 'normal' with $F(\mathbf{r,r^{\prime }})=0$. The carries could be normal-state fermions, as in any BCS-like theory of cuprates, normal-state charged bosons, as in the bipolaron theory \cite{alebook}, or a mixture of both.

 This work was supported by the Leverhulme Trust (grant F/00261/H).

\end{document}